\documentclass{emulateapj}

\slugcomment{Submitted to {\it The Astrophysical Journal}}
\shorttitle{Follow-up Studies of SDSS J1426+5752} 
\shortauthors{E.M. Green, P. Dufour, G. Fontaine, \& P. Brassard}

\newcommand{\gta}{\lower 0.5ex\hbox{$ \buildrel>\over\sim\ $}}
\newcommand{\lta}{\lower 0.5ex\hbox{$ \buildrel<\over\sim\ $}}
\newcommand{\Teff}{T_{\rm eff}}
\newcommand{\Msun}{M_{\rm \odot}}

\begin{document}

\title{Follow-up Studies of the Pulsating Magnetic White Dwarf SDSS
  J142625.71+575218.3 }

\author{E.M. Green\altaffilmark{1},P. Dufour\altaffilmark{1,2},
  G. Fontaine\altaffilmark{2},P. Brassard\altaffilmark{2}}

\altaffiltext{1}{Steward Observatory, University of Arizona, 933 North
  Cherry Avenue, Tucson, AZ 85721; bgreen@as.arizona.edu}
\altaffiltext{2}{D\'epartement de Physique, Universit\'e de
  Montr\'eal, Montr\'eal, QC H3C 3J7, Canada;
  dufourpa@astro.umontreal.ca, fontaine@astro.umontreal.ca,
  brassard@astro.umontreal.ca}

\begin{abstract}

We present a follow-up analysis of the unique magnetic luminosity-variable
carbon-atmosphere white dwarf SDSS J142625.71+575218.3. This includes
the results of some 106.4 h of integrated light photometry which have
revealed, among other things, the presence of a new periodicity at
319.720 s which is not harmonically related to the dominant oscillation
(417.707 s) previously known in that star. Using our photometry and
available spectroscopy, we consider the suggestion made by
Montgomery et al. (2008) that the luminosity variations in SDSS
J142625.71+575218.3 may not be caused by pulsational instabilities, but
rather by photometric activity in a carbon-transferring analog of AM
CVn. This includes a detailed search for possible radial velocity
variations due to rapid orbital motion on the basis of MMT spectroscopy. 
At the end of the exercise, we unequivocally rule out the interacting binary
hypothesis and conclude instead that, indeed, the luminosity variations
are caused by $g$-mode pulsations as in other pulsating white dwarfs.
This is in line with the preferred possibility put forward by Montgomery
et al. (2008).

\end{abstract}


\section{INTRODUCTION}

The rather faint ($g = 19.16$) star SDSS J142625.71+575218.3 (referred
to hereafter as SDSS J1426+5752) is a fascinating object in several
aspects. First, it belongs to the newly-discovered type of
carbon-atmosphere white dwarfs, also known as Hot DQ stars (Dufour et
al. 2007, 2008a). These are exceedingly rare stars whose unexpected
existence was revealed thanks to the availability of some of the data
products that came out of the Sloan Digital Sky Survey (e.g., Liebert et
al. 2003 and Eisenstein et al. 2006). Dufour et al. (2008b) found only
nine such objects out of a total of about 10,000 white dwarfs identified
spectroscopically. Their preliminary atmospheric analysis revealed that
all the Hot DQ white dwarfs fall in a narrow range of effective
temperature, between about 18,000 and 24,000 K, and that they have
atmospheric carbon-to-helium number ratios ranging from 1 to upward of
100. Dufour et al. suggested that these stars could be the 
cooled-down versions of the, so far, unique and very hot ($\Teff$
$\simeq$ 200,000 K) carbon-rich PG 1159 star H1504 (see, e.g., Werner \&
Herwig 2006) and form a new family of hydrogen- and helium-deficient
objects following the post-AGB phase. In this scenario, residual helium
would float rapidly to the surface after the PG 1159 phase of evolution,
and the descendants of H1504-like stars would thus ``disguise''
themselves as helium-atmosphere white dwarfs (of the DO and  DB spectral 
types). This would last until convective mixing dilutes the thin
outermost layer of helium in the effective temperature range where
substantial subphotospheric convection due to carbon recombination
develops in models of these stars. Hence, a dramatic change in the
atmospheres of such stars, from helium-dominated to carbon-dominated,
would occur in the range of temperature where the Hot DQ's are actually
found. Further evolution would slowly restore the dominance of helium in
the atmosphere of these objects as a result of diffusion. Although quite
a bit of work needs to be done to establish quantitatively the
foundations of this scenario, the preliminary investigations of Althaus
et al. (2009) indicate that it is quite viable. An updated discussion of
the properties of Hot DQ stars has been presented by Dufour et al. (2009).

The second interesting development concerning SDSS J1426+5752 was the
important discovery by Montgomery et al. (2008) that it is a luminosity
variable. On the basis of 7.8 h of integrated light photometry on the
McDonald Observatory 2.1 m Otto Struve Telescope, these authors reported
that SDSS J1426+5752 has a light curve dominated by a single periodicity
at 417.7 s with an amplitude of about 1.7\% of the mean brightness of
the star, accompanied by its first harmonic (208.9 s) with a relatively
large amplitude ($\sim$0.7\%), and possibly also by its fourth harmonic
as well ($\sim$0.3\%). Quite interestingly, they also reported that no
luminosity variations were detected in five other Hot DQ's that they
surveyed. Using some theoretical arguments, Montgomery et al. (2008)
argued that the luminosity variations seen in SDSS J1426+5752 and not in
their other targets could be accounted for naturally in terms of pulsational
instabilities. If true, this means that SDSS J1426+5752 is the prototype
of a new class of pulsating white dwarfs after the GW Vir, V777 Her, and
ZZ Ceti types (and see, e.g., Fontaine \& Brassard 2008 for a detailed
review on these pulsators). The hypothesis that the luminosity
variations seen in SDSS J1426+5752 are caused by pulsational
instabilities associated with low-order and low-degree gravity-mode
oscillations (as in the known types of pulsating white dwarfs) is backed
by the exploratory nonadiabatic calculations carried out independently
by Fontaine, Brassard, \& Dufour (2008) in parallel to the efforts of 
Montgomery et al. (2008). 

On the other hand, Montgomery et al. (2008) also noted that the folded
light curve of SDSS J1426+5752 does not resemble those of pulsating
white dwarfs showing nonlinearities in their light curves, but shows
instead similarities with the folded pulse shape of AM CVn, the
prototype of the family of helium-transferring cataclysmic variables. 
The AM CVn stars are close interacting binaries consisting of (probably)
two helium white dwarfs with orbital periods in the range 1000$-$3000 s
(and see the reviews of Warner 1995 or Nelemans 2005 for a lot more
details on these challenging objects). In these systems, the main
photometric period, almost always accompanied by several harmonics,
corresponds to the beat period between the orbital period and the
precession period of the slightly elliptical accretion disk around the
more massive white dwarf. The dominant component of the light
variability usually comes from the moving (precessing) optically thick
accretion disk. Thus, on the basis of similarities in the folded light
pulses between SDSS J1426+5752 and AM CVn, Montgomery et al. (2008)
proposed an alternative to pulsational instabilities for explaining its
luminosity variations: the possibility that it is, in fact, a new type
of close interacting binary, a carbon-transferring analog of AM CVn. In
this scenario, the observed spectrum of SDSS J1426+5752 would originate
from an optically thick carbon-oxygen accretion disk around the more
massive white dwarf component in the system. The pulse shape argument
was again used recently by Barlow et al. (2008) to favor the close
interacting binary model after those other authors discovered two more
luminosity variable Hot DQ's. However, counterarguments, favoring this
time the pulsation model, have been put forward by Dufour et al. (2009)
and Fontaine et al. (2009). 

The third development concerning SDSS J1426+5752 resulted from follow-up
spectroscopic observations carried out by Dufour et al. (2008c) at the
6.5 m Multiple Mirror Telescope (MMT) and at one of the 10 m Keck
Telescopes. This was motivated by the sole availability of a rather poor
SDSS spectrum, insufficient for quantitative analysis. Their objective
was, firstly, to obtain a sufficiently good spectrum for detailed
atmospheric modeling and, secondly, to search for the presence of
helium as required to account for the observed pulsational
instabilities according to the nonadiabatic calculations of Fontaine et
al. (2008). The spectral analysis of the improved spectra readily
revealed the presence of a substantial amount of helium in the
atmosphere of SDSS J1426+5752, an abundance comparable to that of
carbon. This is in line with the expectations of nonadiabatic 
pulsation theory which requires an important ``helium pollution'' in the
atmosphere/envelope of SDSS J1426+5752 for it to pulsate at its current
estimated effective temperature and surface gravity (Dufour et
al. 2008b). In addition to this finding, an unexpected surprise came out
of the follow-up spectroscopic observations of Dufour et
al. (2008c). Indeed, it was found that the strong carbon lines seen in
the optical spectrum of SDSS J1426+5752 feature Zeeman splitting, a
structure that could not be seen in the original noisy SDSS
spectrum. The observed splitting between the $\pi$ and $\sigma$
components implies a large scale magnetic field of about 1.2 MG. Hence,
if SDSS J1426+5752 is really a pulsating star, it would be the first
example of an isolated pulsating white dwarf with a large detectable
magnetic field. As such, it would be the white dwarf equivalent of a
rapidly oscillating Ap (roAp) star. The roAp stars are main sequence (or
near main sequence) magnetic A stars (Ap) showing multiperiodic
luminosity variations with periods in the range 5$-$15 minutes caused by
low-degree, high-order pressure-mode pulsational instabilities (see,
e.g., Kurtz 1990).  

In view of the importance of SDSS J1426+5752, we carried out additional
studies of that star. In particular, we present and discuss in this
paper the results of some 106.4 h of integrated light photometry
gathered at the Steward Observatory 1.55 m Kuiper Telescope. We also 
discuss the pros and the cons of the pulsational instabilities model
versus those of the interacting binary model. This includes a detailed
search for possible radial velocity variations due to rapid orbital
motion using MMT spectroscopy. At the end of the exercise, we
unequivocally conclude in favor of the pulsation model.

\section{TIME-SERIES PHOTOMETRY}

\subsection{Observations}

Our follow-up photometry was obtained with Mont4K (as in ``Montr\'eal
4K$\times$4K camera''), a new CCD camera designed and built at the 
Steward Observatory. Mont4K is a partnership between Universit\'e de
Montr\'eal and the University of Arizona. The instrument was designed
primarily with differential time-series photometry in mind (in the 
windowing mode), but the variety of filters available coupled with the
excellent sensitivity of the chip and the large
(9.7$^{\prime}$$\times$9.7$^{\prime}$) field of view make it ideal for
many imaging projects. It is used at the Kuiper Telescope on Mount
Bigelow near Tucson. Some details about the instrument can be found in
Randall et al. (2007), but the interested reader will find more on the
following web site:
http://james.as.arizona.edu/~psmith/61inch/instrument s.html.   

At the outset, the relative faintness of SDSS J1426+5752 posed a
challenge for a small instrument such as the Kuiper Telescope, but we
were pleasantly surprised at the quality of the light curves that we
could actually obtain. The time-series observations were taken on dark nights
through a broadband Schott 8612 filter, and an effective exposure time
of 67 s (on average) was used as a compromise between the S/N and
the need to sample adequately the luminosity variations as reported by
Montgomery et al. (2008). Altogether, we collected some 106.4 h of useful 
photometry over a 40 day span in the spring of 2008. This corresponds to a
formal temporal resolution of 0.29 $\mu$Hz and a modest duty cycle of
11.8\%.  Details of the observations are provided in Table 1.

% Table 1
\begin{deluxetable}{cccc}
\tablewidth{0pt}
\tablecaption{Journal of Observations for SDSS J1426+5752}
\tablehead{
\colhead{Date} &
\colhead{Start of Run} &
\colhead{Number of Frames} &
\colhead{Length}\\
\colhead{(UT)} &
\colhead{(HJD2454550+)} &
\colhead{} &
\colhead{(h)}
}
\startdata
2008 Mar 31 & 6.6517743 & 399 & 7.88\\
2008 Apr 01 & 7.6417324 & 468 & 8.73\\
2008 Apr 03 & 9.6413232 & 466 & 8.69\\
2008 Apr 04 & 10.6382536 & 449 & 8.52\\
2008 Apr 05 & 11.6310329 & 223 & 4.15\\
2008 Apr 06 & 12.6288716 & 487 & 9.05\\
2008 Apr 09 & 15.6271095 & 364 & 6.55\\
2008 Apr 10 & 16.6864942 & 108 & 1.99\\
2008 Apr 11 & 17.7342303 & 328 & 6.11\\
2008 May 01 & 37.6670997 & 389 & 7.26\\
2008 May 02 & 38.6404921 & 424 & 7.88\\
2008 May 03 & 39.6412906 & 418 & 7.80\\
2008 May 04 & 40.6406385 & 275 & 6.29\\
2008 May 08 & 44.6459373 & 406 & 7.65\\
2008 May 10 & 46.6446166 & 420 & 7.82\\
\enddata
\end{deluxetable}

Figure 1 shows the 15 nightly light curves that were obtained. The
original images were reduced using standard IRAF reduction aperture
photometry routines, except that we set the photometric aperture size
individually for each frame to 2.0 times the FWHM in that image. We
computed differential light curves of SDSS J1426+5752 on the basis of
three suitable comparison stars well distributed around the
target. Final detrending of the effects of differential extinction was
made through a spline strategy. A zoomed-in view of the light curve
gathered on May 2 (2008) is provided in Figure 2. Again, in view of the
magnitude of the target ($g$ = 19.16), the overall quality of the light
curves is most gratifying. We attribute this to the excellent
sensitivity of the CCD and our optimized data pipeline.

\begin{figure}[!ht]
\plotone{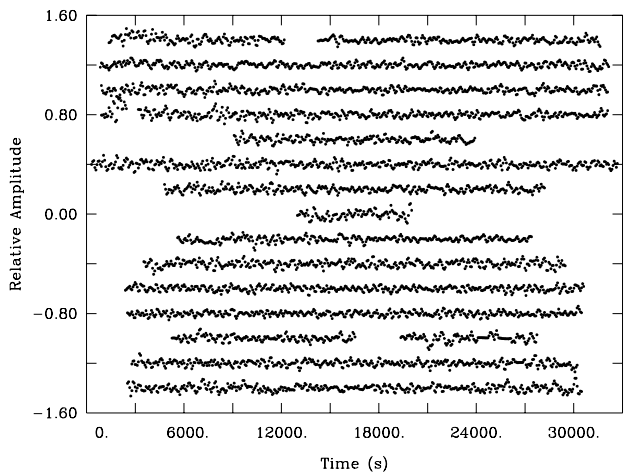}
\caption{All light curves obtained for SDSS J1426+5752
using the Mont4K CCD camera mounted on the Steward Observatory 1.55 m 
Kuiper telescope. The data have been shifted arbitrarily along the x and
y axes for visualization purposes. They are expressed in units of
fractional brightness intensity and seconds. From top to bottom, the
curves refer to the nights of March 31, April 1, April 3, April 4,
April 5, April 6, April 9, April 10, April 11, May 1, May 2, May 3,
May 4, May 8, and May 10 (2008). For details, see Table 1.}
\end{figure}

\begin{figure}[!ht]
\plotone{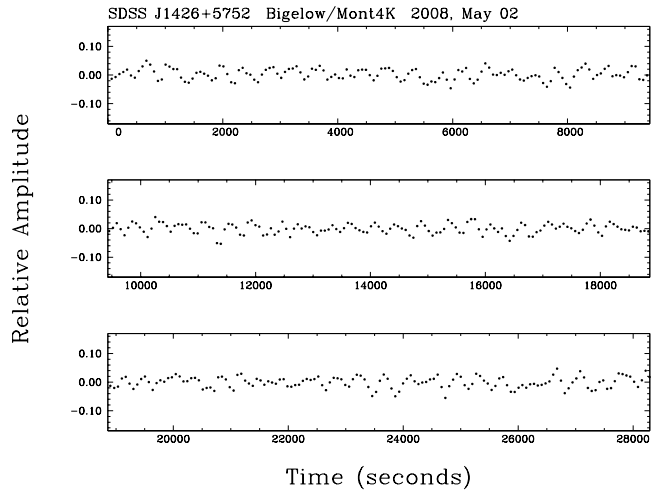}
\caption{Expanded view of the light curve of SDSS
J1426+5752 obtained on May 2, 2008. The units are the same as those used
in Fig. 1. There is a hint that the minima are generally sharper than the
maxima.}
\end{figure}

\subsection{Frequency Analysis}

The time-series photometry gathered for SDSS J1426+5752 was analyzed in a
standard way using a combination of Fourier analysis, least-squares fits
to the light curve, and prewhitening techniques (see, e.g., Bill\`eres
et al. 2000 for more details). Figure 3 shows the Fourier amplitude
spectrum of the full data set in the 0$-$7.4 mHz bandpass (upper curve)
and the resulting transforms after prewhitening of two (middle curve)
and three significant peaks (lower curve). These results first
confirm the presence of a dominant oscillation with a period of 417.707
s and of its first harmonic at 208.853 s, in agreement with the report
of Montgomery et al. (2008). Note, in this context, that we did not
detect the 2nd harmonic, again in agreement with their report, but also
that we could not verify the presence of the 4th harmonic of the main
periodicity as suggested in the data of Montgomery et al. (2008) because
our relatively large sampling time led to an effective Nyquist frequency
of 7.46 mHz, too small for picking up that high-frequency peak. On the
other hand, our results also reveal the presence of an additional significant
oscillation with a period of 319.720 s as can be seen in the figure. 
It is also quite likely that there are many other frequency components
in the light curve of SDSS J1426+5752, independent oscillations or
rotationally-split components for instance, but the sensitivity of our
observations did not allow them to be detected.

\begin{figure}[!ht]
\plotone{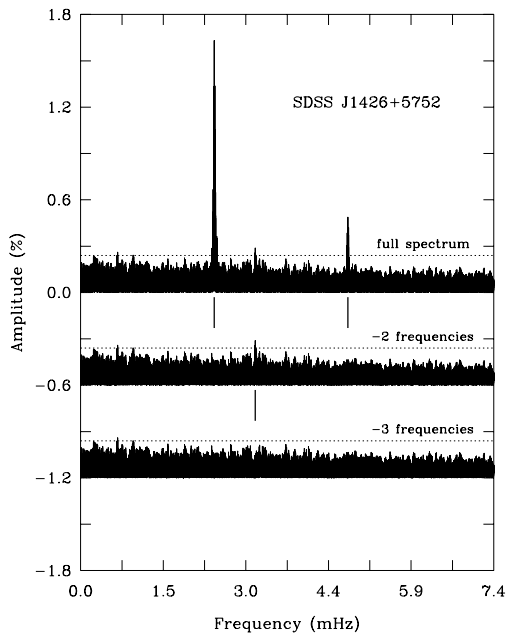}
\caption{Fourier transform of the entire data set in the
0$-$7.4 mHz range. The lower transforms show the successive steps of
prewhitening by the two strongest frequencies (the 417.707 s peak and
its first harmonic), and finally by all three frequencies with
statistically significant amplitudes. The dotted horizontal lines
indicate the 4 $\sigma$ noise level.}
\end{figure}

The two upper curves in Figure 4 provide a zoomed in view of the Fourier
amplitude spectrum in the vicinity of the 417.707 s peak before and
after prewhitening. Likewise, the two lower curves give a similar view
for the 319.720 s peak. It is interesting to note that, within our
measurement errors, the main peak at 417.707 s is a singlet. Given that
SDSS J1426+5752 has a large scale magnetic field of some 1.2 MG, one
could have expected instead the presence of $l+1$ components due to
magnetic splitting according to Jones et al. (1989). Perhaps the
multiplet components have much lower amplitudes than the main mode and
are buried in the noise, or perhaps the field geometry is such that
magnetic splitting cannot be observed. Either way, this remains an
interesting curiosity.

% Table 2
\begin{deluxetable*}{cccc}
\tablewidth{0pt}
\tablecaption{Harmonic Oscillations Detected in the Light Curve of SDSS
  J1426+5752 }
\tablehead{
\colhead{Period} &
\colhead{Frequency} &
\colhead{Amplitude} &
\colhead{Phase}\\
\colhead{(s)} &
\colhead{(mHz)} &
\colhead{(\%)} &
\colhead{(s)}
}
\startdata
417.70687$\pm$0.00082 & 2.394023$\pm$0.000005 & 1.630$\pm$0.048 &
127.90$\pm$1.97\\
319.72035$\pm$0.00271 & 3.127733$\pm$0.000026 & 0.288$\pm$0.048 &
239.89$\pm$8.54\\
208.85344$\pm$0.00068 & 4.788047$\pm$0.000015 & 0.487$\pm$0.048 &
21.06$\pm$3.30\\
\enddata
\end{deluxetable*}

\begin{figure}[!ht]
\plotone{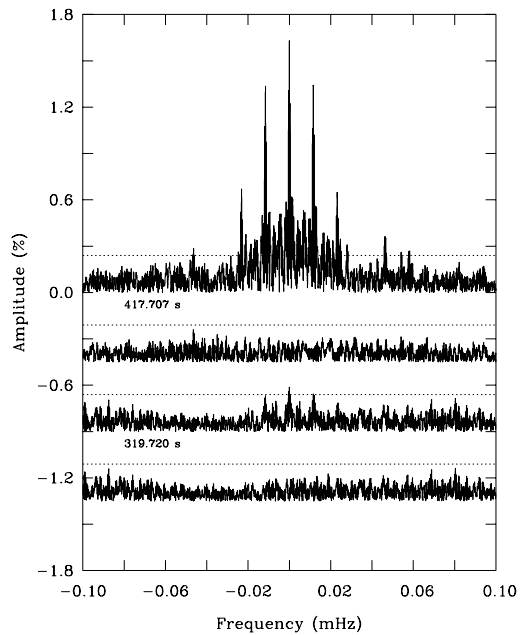}
\caption{Zoomed in view of the Fourier transform of the
entire data set in the vicinity of the 417.707 s peak (the two upper
curves) and of the 319.720 s peak (the two lower curves). The two curves
asssociated with a given oscillation show the transform before and after
prewhitening. The dotted horizontal lines indicate the 4 $\sigma$ noise
level.}
\end{figure}

We thus were able to extract three distinct harmonic oscillations in the
light curve of SDSS J1426+5752. The basic characteristics of these
oscillations are summarized in Table 2. Note that the phase is relative
to an arbitrary point in time; in our case, the beginning of the first
run on UT 31 March 2008. The uncertainties on the period, frequency,
amplitude, and phase of each oscillation as listed in the table were
estimated with the method put forward by Montgomery \& O'Donoghue
(1999). We point out, in this context, that the uncertainties on the
amplitudes and phases obtained by our least-squares fits during the
prewhitening stage were virtually the same as those derived with the
Montgomery \& O'Donoghue (1999) method. A basic quantity in that latter
approach is the average noise level in the bandpass of interest, and we
thus computed this mean value from the residual Fourier transform (the
lower curve in Fig. 3) spanning the 0$-$7.4 mHz interval. The mean noise
level in that range turned out to be 0.060\% of the mean brightness of
the star. In comparison, if we use the shuffling technique discussed by
Kepler (1993) for estimating the noise level, we find an almost
identical value of 0.061\%.

In the context of being able to discriminate between the two
possibilities put forward by Montgomery et al. (2008) to explain the
luminosity variations of SDSS J1426+5752 -- pulsations versus
interacting binary -- our detection of a periodicity (319.720 s) that is
incommensurate with the periods of the main peak and its harmonics is
potentially quite important (see below). Hence, it is essential that its
presence be well established. Formally, according to the results of
Table 2, our detection of the 319.720 s periodicity is a 6.0 $\sigma$
(0.288/0.048) result. In the more standard way, the derived amplitude is
rather compared to the average noise level, in which case our detection
would be seen as a 4.8 $\sigma$ (0.288/0.060) result. If we divide our
runs into two ``seasons'', i.e., from March 31 through April 11, and
then from May 1 through May 10, we find that the 319.720 s period is
present in both sets of observations (319.715$\pm$0.011 s and
319.738$\pm$0.013 s), at the level of 4.2 $\sigma$ according to the
standard criterion. We also verified explicitly that the 319.720 s
periodicity is not present in the light curves of our comparison stars,
thus ruling out a possible instrumental effect. We note finally that
Montgomery et  al. (2008) would not have been able to pick up the
319.720 s periodicity in their data -- assuming that it was present in
their light curve with an amplitude comparable to our detection --
because their sensitivity was at least a factor of two lower than what
we could achieve on a smaller telescope (1.55 m versus 2.1 m) at a
brighter site but at the price of much longer observations (106.4 h
versus 7.8 h). 

\subsection{Amplitude and Phase Variations}

We investigated the stability of the amplitude and phase of each of the
three frequencies we extracted from the light curve of SDSS J1426+5752
by performing nightly measurements. A clue about possible amplitude
variations on a daily timescale is first provided by Figure 5 in which we
show a montage of the nightly Fourier amplitude spectra. Considering the
main peak (2.3950 mHz, 417.707 s), the figure does suggest some possible
amplitude variations. However, things are a lot less suggestive for the
two other peaks given the level of noise and their relatively small
amplitudes.  

\begin{figure}[!ht]
\plotone{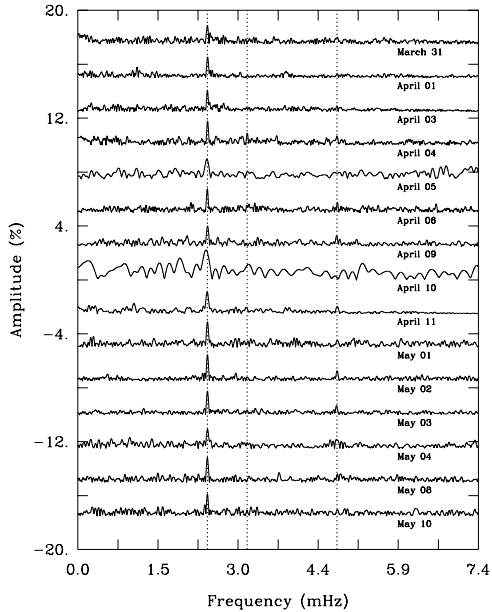}
\caption{Montage of the nightly Fourier transforms, shifted
arbitrarily along the y-axis for visualization purposes. The locations
of the three detected frequencies are indicated by the vertical dotted
lines. }
\end{figure}

A more quantitative and standard way of measuring the nightly amplitudes
(and phases) is to fix the periods at their values given in Table 2 and
simultaneously perform least-squares sine fits with these periods for each
nightly run. The output are nightly amplitudes and phases with formal
estimates of their uncertainties. It is interesting to point out
that the formal estimates of the uncertainties on the amplitudes and
phases that came out of our least-squares exercise were, again,
essentially the same as those obtained through the method of Montgomery
\& O'Donoghue (1999), which we explicitly used after the fact as a
verification. 

Figure 6 summarizes our results in the case of the main periodicity
found in the light curve of SDSS J1426+5752. Note that we explicitly
excluded the values obtained from the short runs gathered on April 5 and
April 10 because the associated uncertainties are significantly larger
than those shown in the figure. Hence, the top panel in Figure 6
displays the amplitudes of the 417.707 s peak along with their
formal 1 $\sigma$ uncertainties for 13 nightly runs. The central dotted
horizontal line represents the weighted average of the nightly
amplitudes and the horizontal lines above and below the average value
give the 1 $\sigma$ uncertainty on that value. In the case of the phase
(lower panel of Fig. 6), a similar procedure was used, except that the
average value was shifted to zero (since the phase is arbitrary) and
ultimately expressed in units of cycle. 

\begin{figure}[!ht]
\plotone{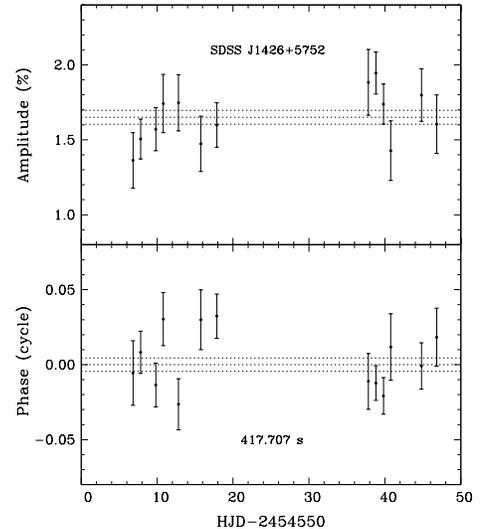}
\caption{Nightly measurements of the amplitude and phase of
the 417.707 s harmonic oscillation seen in SDSS J1426+5752. All nights
were included, except those of April 5 and April 10 corresponding to
rather short runs (see Table 1) and, consequently, to nightly estimates
with uncertainties significantly larger than those typically illustrated
here. }
\end{figure}

Taken at face value, the results summarized in Figure 6 do suggest some
variations in amplitude and in phase over timescales of days. For
instance, on March 31, we find a ``low'' amplitude of 1.36$\pm$0.18\%  
for the dominant periodicity at 417.707 s, while we find a ``high''
amplitude of 1.95$\pm$0.14\% on May 2. In this context, it may
be appropriate to recall that amplitude modulation associated with the
rotation of the star is typically seen in roAp stars (see, e.g., Kurtz
1990). This phenomenon (and many others) observed in roAp stars has been
explained within the framework of the very successful oblique pulsator
model. In that model, the pulsations align themselves along the magnetic
field axis which is itself inclined with respect to the rotation axis of
the star. The viewing aspect of the pulsations thus changes periodically
with rotation, which produces amplitude modulation of a given mode. In
the case of SDSS J1426+5752, we do not yet know if its large scale
magnetic field is aligned with the rotation axis.

Although the material presented in this subsection is suggestive of
possible amplitude and phase modulations with timescales of days for the
dominant 417.707 s period, it is simply not possible at this stage to be
certain about their reality. If, for example, we were to double the
uncertainties on the derived nightly amplitudes and phases (assuming that our
least-squares approach or the method of Montgomery \& O'Donoghue 
1999 underestimates the true errors by that factor), then we could only
conclude that the nightly amplitudes and phases of the 417.707 s
periodicity we  extracted from the light curve of SDSS J1426+5752 do not
vary within our measurement errors. Things are even worse for the lower
amplitude 319.720 s and 208.853 s oscillations in that the uncertainties
on their  nightly amplitudes and phases prevent us from concluding with any
certainty about possible modulations although there are hints of
variations. Here then is a classic case of ``more observations
are needed''.

\subsection{Pulse Shape}

We have followed up on the remark made by Montgomery et al. (2008) that
the folded pulse shape of SDSS J1426+5752 is different from that of 
pulsating white dwarfs and rather shows similarities with that of AM CVn, 
the prototype of helium-transferring double degenerate  binaries. The
top panel of Figure 7 shows our 106.4 h long light curve of SDSS 
J1426+5752 folded on the period of 417.707 s. To reach a decent S/N, we
distributed the folded amplitudes in 10 different phase bins, each containing
572 points on average. The error bars about each point in the folded
light curve correspond to the errors of the mean in each bin. Given that 
the first harmonic of the 417.707 s periodicity in the light curve has a
very high amplitude, about 30\%  of that of the main peak (see Table 2),
it is not surprising that the pulse shape illustrated in the top panel
of Figure 7 is highly nonlinear. Along with this, another striking 
characteristic with respect to known pulsating white dwarfs is
the fact that it boosts a relatively flat maximum and a sharp minimum
(see Montgomery et al. 2008 and the examples below).

\begin{figure}[!ht]
\plotone{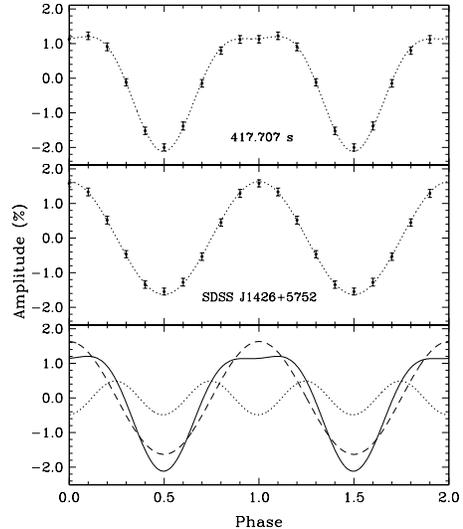}
\caption{{\it Top panel:} Light curve of SDSS J1426+5752 folded
on the period of 417.707 s and distributed in 10 phase bins (points with
error bars). The dotted curve is a model pulse shape obtained by summing
two sinusoids with the periods, amplitudes, and phases of the 417.707 s
periodicity and of its first harmonic (208.853 s) as listed in Table 2. 
{\it Middle panel:} Light curve of SDSS J1426+5752 folded on the period of
417.707 s and distributed in 10 phase bins after prewhitening of the
first harmonic component. The dotted curve is a pure sine wave computed
using the period, amplitude, and phase of the dominant 417.707 s
periodicity. {\it Bottom panel:} The dotted curve is a pure sinusoid
computed with the period, amplitude, and phase of the 208.853 s
harmonic; the dashed curve is a pure sine wave computed
using the period, amplitude, and phase of the dominant 417.707 s
periodicity (this is the same as the dotted curve in the middle panel);
the solid curve is the sum of these two sinusoids and corresponds to the
model pulse shape represented by the dotted curve in the top panel. }
\end{figure}

In the middle panel of Figure 7, we again display our folded light
curve of SDSS J1426+5752, but only after having prewhitened the data of
the first harmonic (208.853 s) of the main periodicity. If higher order
harmonics have negligible amplitudes, and if other modes do not
interfere in the folding process\footnote{We explicitly verified that the
results displayed in both the upper and middle panel of Fig. 7 are
completely insensitive to whether or not the data is prewhitened of the
low-amplitude 319.720 s oscillation. This should not be surprising given
the large number of 319 s cycles in 106.4 h, but for very short runs,
prewhitening of the other modes is required in this sort of exercise},
the pulse shape in the middle panel should be that of a perfect sinusoid
with an amplitude equal to that of the 417.707 s oscillation described
in Table 2. This is precisely what the dotted curve shows. Taking into account
the uncertainties, the match between the folded pulse shape and the
template is essentially perfect.

In the lower panel, we plotted the same template (dashed curve)
corresponding to the 417.707 s sinusoid. Taking properly into account
the phase difference, we also plotted a sinusoid (dotted curve) with the
defining characteristics of the 208.853 s periodicity as given in Table
2. The sum of these two sine waves gives the solid curve, the overall
nonlinear pulse shape associated with the 417.707 s oscillation. We
reported this model pulse shape in the upper panel of Figure 7 (now as a
dotted curve) so that a direct comparison can be made with the
observations. Again, within our measurement errors, the agreement is
nearly perfect. One can see that the relatively flat maximum and sharp
minimum in the pulse shape is due to the fact that the first harmonic of
the main oscillation falls nearly in phase at the minimum and in
antiphase near the maximum of the main sinusoid. This is particularly
well illustrated in the lower panel of the figure. 

For comparison purposes, we carried out similar folding exercises using
representative light curves from the archives that one of us (G.F.)
built up over the years using LAPOUNE at the 3.6 m Canada-France-Hawaii
Telescope (CFHT). LAPOUNE is a portable three-channel photometer that uses
photomultiplier tubes as detectors. The archived light curves are
integrated ``white light'' data and were generally obtained with a
sampling time of 10 s. They include a short (2.04 h) light curve of AM
CVn itself, originally taken as a test as part of a multisite campaign
carried out on that star (Provencal et al. 1995; Solheim et
al. 1998). In a format identical to that of Figure 7, Figure 8
summarizes the results of our calculations after having folded that 
light curve of AM CVn on the period of 521.105 s, which is the dominant
photometric periodicity in that star (see Provencal et al. 1995). Although 
the model pulse shape (dotted curve in upper panel and solid curve in
lower panel) based on the superposition of two sinusoids (the 521.105 s
oscillation and its first harmonic) is far from perfect, it is
nevertheless sufficient for illustrating the fact that AM CVn does show
a main pulse shape with a rounded top and a sharp bottom. The lower
panel of Figure 8 explains why: the first harmonic (dotted curve) of
the 521.105 s oscillation shows a relatively large amplitude compared to
that of the main component, and it tends to be in phase and in antiphase
at the minimum and maximum, respectively, of that component. There is a
significant offset however, between the extrema of the 521.105 s
sinusoid and those of its first harmonic, and this largely explains the
asymmetric shape of the pulse in this particular light curve.

\begin{figure}[!ht]
\plotone{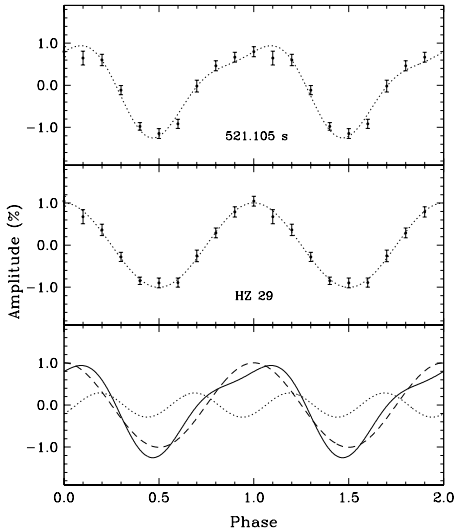}
\caption{Similar to Fig. 7, but referring to AM CVn ($V$ =
14.1$-$14.2), also known as HZ 29, the prototype of helium-transferring
double degenerate  binaries. The light curve was folded on the period
of 521.105 s and the first harmonic of that periodicity is also involved
in the plot. The light curve comes from a single run, 2.04 h long,
gathered in white light using the CFHT/LAPOUNE combination. The folded
light curve has been distributed in 10 phase bins, each
containing $\sim$74 points. }
\end{figure}

In contrast to this behavior, known pulsating white dwarfs with
nonlinear light curves (the large amplitude ones) quite generally
display pulse shapes that have flatter minima and sharper maxima. An
example of this is provided by Figure 9 which refers to the large
amplitude ZZ Ceti star PG 2303+242, a pulsator with a light curve
dominated by a periodicity at 712.250 s. In the present case, the model
pulse shape based on only two components (the dominant mode and its
first harmonic) does a very good job at explaining the folded light
curve, as can be appreciated in the top panel of Figure 9.

\begin{figure}[!ht]
\plotone{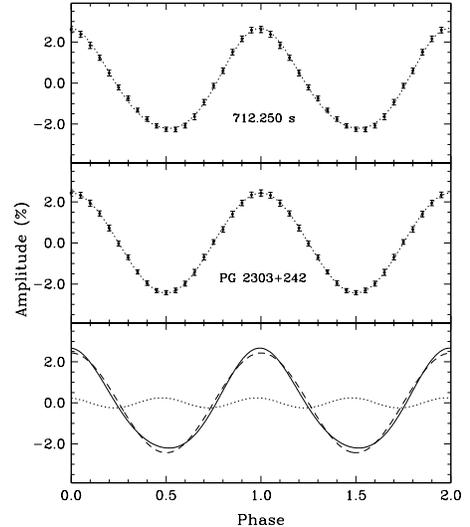}
\caption{Similar to Fig. 7, but referring to the ZZ Ceti
star PG 2303+242 ($V$ = 15.50). The light curve was folded on the period
of 721.250 s and the first harmonic of that periodicity is also involved
in the plot. The light curve comes from a single run, 4.25 h long,
gathered in white light using the CFHT/LAPOUNE combination. The folded
light curve has been distributed in 20 phase bins, each
containing $\sim$77 points. }
\end{figure}

A second example is provided by another large amplitude ZZ Ceti star, GD
154, as illustrated in Figure 10. In that case, we retained also the
second harmonic along with the main oscillation (1186.085 s) and its
first harmonic in the construction of the model pulse shape. Although
the top panel of the figure clearly demonstrates that the model could be
improved (GD 154 has a complicated multiperiodic light curve with
higher-order harmonic terms; see, e.g., Fig. 23 of Fontaine \& Brassard
2008), the results presented here are sufficient to make our point about
the general shape of a light pulse in a large amplitude pulsating white
dwarf.

\begin{figure}[!ht]
\plotone{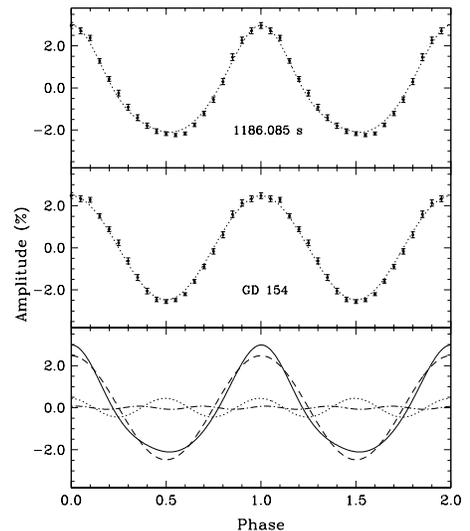}
\caption{Similar to Fig. 7, but referring to the ZZ Ceti
star GD 154 ($V$ = 15.33). The light curve was folded on the period
of 1186.085 s and, this time, $both$ the first harmonic (dotted curve in
the bottom panel) $and$ the second harmonic (dot-dashed curve in the bottom
panel) of that periodicity are also involved in the plot. The folded
light curve in the middle panel has been obtained after prewhitening of
the first and second harmonic contributions. The light curve comes from
two consecutive nights covering a length of 8.76 h and was gathered in
white light using the CFHT/LAPOUNE combination. The folded light curve
has been distributed in 20 phase bins, each containing $\sim$158
points. }
\end{figure}

Another interesting example is shown in Figure 11. It concerns Balloon
090100001, which is not a pulsating white dwarf at all, but instead
belongs to the family of hot B subdwarf pulsators (and see Fontaine et
al. 2006 for a brief review on these stars, if interested). The light
curve of that star, the largest amplitude variable of that type currently
known, is dominated by a main pulsation (the fundamental radial mode 
according to Van Grootel et al. 2008) with a period of 356.194
s. Our model pulse shape, based on the superposition of the main
periodicity and of its first two harmonics, gives a perfect fit to
the folded light curve.

\begin{figure}[!ht]
\plotone{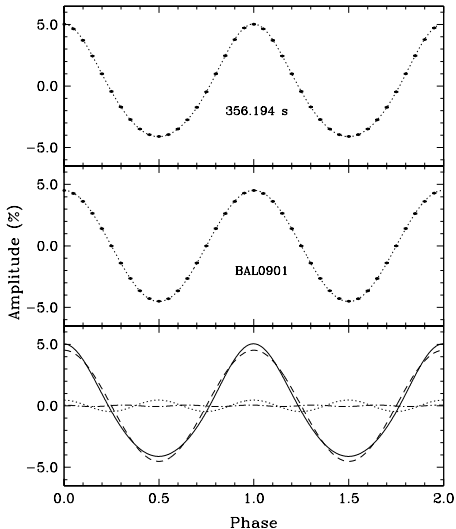}
\caption{Similar to Fig. 10, but referring to the
pulsating hot B subdwarf star Balloon 090100001 ($V$ = 12.10). The light
curve was folded on the period of 356.194 s and both lowest harmonics of 
that periodicity are also involved in the plot. The light curve comes from
four consecutive nights covering a length of 17.81 h and was gathered in
white light using the CFHT/LAPOUNE combination. The folded light curve
has been distributed in 20 phase bins, each containing $\sim$321 points.}
\end{figure}

What is common in these examples, and what is generally true for large
amplitude pulsating white warfs and hot subdwarfs, is that the harmonic
components tend to be in phase at light maximum. This produces
relatively sharp maxima and flat minima. In contrast, the first harmonic
tends to be in antiphase with the main periodicity at light maximum in
the light curve of SDSS J1426+5752 and AM CVn, and this now produces
pulse shapes with rounded tops and sharp minima. Montgomery et
al. (2008) interpreted this as evidence that the luminosity variations
observed in SDSS J1426+5752 could be caused, not by pulsational
instabilities (their preferred possibility), but by photometric activity
in a  carbon-transferring analog of AM CVn. We show below, however, that
the pulsation hypothesis provides the better explanation.

\subsection{Pulsations or Interacting Binary?}

On the basis of our photometry, can we argue in favor of one or the
other of the two possibilities put forward by Montgomery et al. (2008) to
account for the luminosity variations seen in SDSS J1426+5752? To begin
with, it is well known that flickering -- incoherent light bursts arising
on short timescales -- is a telltale sign of mass transfer in an
interacting binary system (Warner 1995). Flickering is actually observed in
the CFHT/LAPOUNE light curve of AM CVn as can be readily seen in Figure 1 of
Provencal et al. (1995) where the light curve has been displayed for the
first time. The sampling time was 10 s, sufficiently short to pick up
some flickering. However, when degraded to a sampling time of 60 s,
flickering all but disappears in the light curve of AM CVn. We conclude from
this that we could not have detected flickering in our photometric
observations of SDSS J1426+5752 if present, because of the large sampling
time we used. Hence, we cannot use this as a diagnostic.

On the other hand, the light curves of AM Cvn systems are known to be
unstable and irregular. Solheim et al. (1998) discuss this phenemenon in
some detail in relation to their Figure 2, where the CFHT/LAPOUNE light
curve (binned in 40 s data points) is again displayed. The authors
comment that, in spite of the fact that the CFHT/LAPOUNE light curve is
essentially noise-free, it is irregular compared to that of isolated
pulsating white dwarfs. For instance, the troughs in the light curve are
irregularly spaced in time. Random flickering is blamed for this state of
affairs. In contrast to this, and not withstanding the noise, the light
curves we have gathered (see Fig. 1) have kept the same appearance and
regularity over a six week period. However, it is not clear if this is
really significant since flickering is not coherent over long timebases.

The light curves of AM CVn systems are generally dominated by a
principal periodicity along with a suite of several harmonics of that
dominant oscillation (Warner 1995). Montgomery et al. (2008) reported
the detection of the 417 s oscillation in SDSS J1426+5752 along with its
first harmonic and possibly also the 4th harmonic, but not the 2nd or
3rd one. The detection of a period -- ideally several periods -- that
would be incommensurate with such series of harmonically related
oscillations would go against the interacting binary hypothesis and
would favor the pulsations alternative. This would be the best evidence
for pulsations according to Barlow et al. (2008). Our detection of the
319 s oscillation (see Subsection 2.2) therefore goes a long way in that
direction. 

What about the pulse shape argument? We do not know at this stage why the
folded light curve of SDSS J1426+5752 features a maximum that is flatter
than its minimum, contrary to what is observed in large amplitude pulsating
white dwarfs and hot subdwarfs. However, the fact that the pulse shape
is unusual does not rule out by any means the possibility that
pulsations are involved. In fact, Nature provides us with explicit
examples of isolated pulsating stars with pulse shapes that
qualitatively resemble that seen in SDSS J1426+5752 in that they display
rounded maxima and sharper minima. These are some of the roAp stars with
the largest amplitudes that show harmonics of dominant modes in their
light curves. An excellent case is that of HR 3831 which exhibits a
pulse shape with a rounded top and a sharp bottom as can be seen in
Figure 1 of Kurtz, Shibahashi, \& Goode (1990). The fast rise, rounded
maximum, slower decline, and pointed minimum are very reminescent of
what is observed in the light curve of the AM CVn system CR Boo in its high
state as can be seen in Figure 6 of Warner (1995) for instance. Yet, HR
3831 is a genuinely pulsating star. Another example is provided by the
roAp star HD 99563 studied by Handler et al. (2006). As can be seen in their
Figure 1 (and see also the discussion in the text), the pulsations of
that star show again flatter light maxima than minima. The authors
comment on the phasing of the harmonics with respect to the main mode in
their paper.

SDSS J1426+5752 has one thing in common with roAp stars, and that is a
large scale magnetic field sufficiently important in both cases to
disrupt the atmospheric layers and influence the pulsations there. It
would therefore not surprise us if the magnetic field were 
responsible for the unusual pulse shape (relative to nonmagnetic
pulsating stars) observed in the light curves of SDSS J1426+5752 and
large amplitude roAp stars. Be that as it may, in view of the very
existence of roAp stars such as HR 3831 and HD 99563, the argument that
the pulse shape of SDSS J1426+5752 is different from those seen in
(nonmagnetic) pulsating white dwarfs cannot be held against the
pulsations interpretation.  

\section{SPECTROSCOPY}

\subsection{Search for Radial Velocity Variations}

Dufour et al. (2008c) presented follow-up spectroscopy of SDSS
J1426+5752 including, in particular, an optical spectrum obtained with
the Blue Channel Spectrograph at the MMT using a 500 line mm$^{-1}$
grating with a 1$^{\prime\prime}$ slit, resulting in a $\sim$3.6 \AA~
FWHM spectral resolution over a wavelength range of 3200$-$6400 \AA. A
total of 17 individual spectra were obtained over a timebase of 3.37 h,
resulting in an overall exposure time of 180 minutes. Part of the final
combined spectrum, taken from Dufour et al. (2008c), is reproduced in
Figure 12 (upper curve). It shows a S/N$\sim$75 per pixel at 4500
\AA. The most prominent features are CII lines which clearly show Zeeman
splitting in their cores. As a comparison, we also plotted the available
SDSS spectrum for the star (lower curve).

\begin{figure}[!ht]
\plotone{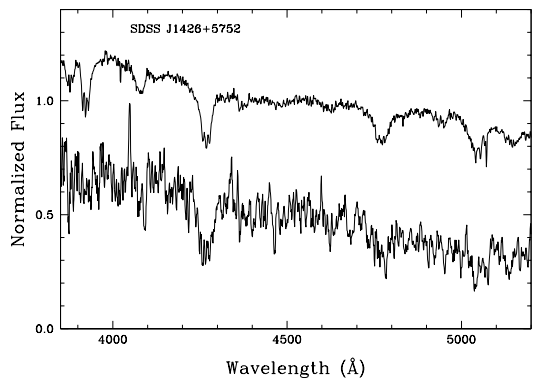}
\caption{Optical spectra of SDSS J1426+5752 obtained 1) at
the MMT (upper curve), and 2) in the SDSS archives (lower curve). Note
that we have applied for clarity a three-point average window smoothing
in the display of the SDSS spectroscopic data.}
\end{figure}

If SDSS J1426+5752 is truly a double degenerate carbon analog of AM CVn
systems, then one should be able to pick up radial velocity variations
associated with orbital periods in the range 1000$-$3000 s as found in
most such systems. This excludes, of course, the improbable configuration of
having the orbital plane at nearly 90 degrees with respect to the
line-of-sight. The sampling and the baseline of 3.37 h used during the
MMT observations of Dufour et al. (2008c) are well suited to search for
such radial velocity variations. In addition, the combination of the
Blue Channel Spectrograph and the MMT is well known to provide a very
stable platform for radial velocity measurements, and one of us (E.M.G.)
has developed over the years a large MMT program dedicated to radial
velocity measurements in hot subdwarf stars. We therefore went back to
the MMT spectroscopic data and searched for possible radial velocity
variations using the tools and expertise developed at Steward
Observatory. We provide some details on the procedure followed.

We first reinterpolated the 17 individual spectra of SDSS J1426+5752
onto a logarithmic wavelength scale with identical starting and ending
wavelengths (3215$-$6357 \AA). After manually removing cosmic ray
signatures from each spectrum, we fitted the continuum, divided by the
fit, and subtracted 1.0 to get a flattened continuum with a mean level
of zero. We median-filtered the 17 resulting spectra into a single
spectrum to create a higher S/N radial velocity template. Radial
velocity cross-correlations of the individual continuum-removed spectra
relative to the template spectrum were performed using the double
precision version of IRAF's FXCOR task, fitting the cross-correlation
peak with a gaussian. A ramp filter was used prior to the
cross-correlation to select the optimal range in Fourier space, with
adopted values of 125, 275, 1480, and 1490, respectively, for the
cut-on, full-on, cut-off, and full-off points. The high frequency noise
cut-off corresponds to 2.75 pixels, which is slightly better than the
instrumental resolution of 3.0 pixels, as determined by the FWHM's of
the HeArNe comparison arc lines. The low-frequency cut-on, corresponding
to 33 pixels, is slightly wider than the observed FWHM (30 pixels) of the
strongest absorption lines in the spectrum. 

The essential result that came out of this exercise is that the
velocities are constant to within 7.1 km s$^{-1}$ (the standard
deviation), which is about 1/32 of the velocity resolution, and there is
no sign of any velocity trend over the 3.37 h of the observations. We
point out that 1/32 of a resolution element represents quite a high
level of accuracy for spectra at this S/N. In comparison, the typical
radial velocity semi-amplitude expected in a AM CVn system is $K = 62$
sin $i$ km s$^{-1}$. This estimate uses the following representative
values: $M_1 = 0.7 \Msun$ for the mass of the helium degenerate primary, 
$M_2 = 0.07 \Msun$ for the mass of the helium degenerate or
semi-degenerate donor, and $P = 2000$ s for the orbital period, as can
be inferred in the reviews of Warner (1995) and Nelemans (2005) and
references therein. Actually, if we assume that the dominant
photometric period of 417 s is approximately equal to the orbital period
in a putative SDSS J1426+5752 carbon system equivalent of AM CVn (this
is the case for the known systems of the type, except for AM CVn
itself), then the estimate of the velocity semi-amplitude goes up to $K
= 105$ sin $i$ km s$^{-1}$. If, as in AM CVn itself, the orbital period
is rather approximately equal to twice the dominant photometric
periodicity of 417 s (521 s in AM CVn), the velocity semi-amplitude
takes on the value $K = 83$ sin $i$ km s$^{-1}$. Hence, unless 
the inclination of the orbital plane is quite low, this argues strongly
against the interacting binary hypothesis as the explanation for the
luminosity variations observed in SDSS J1426+5752.

\subsection{Pulsations or Interacting Binary?}

Along with the fact that we do not detect radial velocity variations,
the available spectroscopic data on SDSS J1426+5752 can further be used to
build up the case against the interacting binary model. In a AM CVn-type
system in its high state, most of the light does not come from the
photosphere of the helium degenerate primary, but rather from an
optically thick accretion disk orbiting around that star. According to
Warner (1995), the HeI lines seen in absorption have quite different
profiles from those observed in isolated helium-atmosphere (DB) white
dwarfs. Their profiles are asymmetric, and they vary both in shape and
depth, contrary to the symmetric and stable lines seen in DB white
dwarfs.

In this context, we wish to point out that the available spectroscopy on
SDSS J1426+5752 shows that the broad (carbon) absorption lines are symmetric
(see, e.g., Fig. 12). Furthermore, the spectrum appears quite stable, at
least over a timescale of 20 h, which is the time lapse between the
observations of SDSS J1426+5752 carried out at the MMT and those
gathered at the Keck I Telescope (and see Dufour et al. 2008c). The
spectrum does not show any sign of accretion disk activity. However, it
does show clear Zeeman splitting in the line cores, a feature that would 
presumably be washed away if SDSS J1426+5752 were part of a interacting
binary system because of the rapid orbital motion. Zeeman splitting is
seen in both the MMT and Keck spectra, and the spacings between the
$\pi$ and $\sigma$ components are the same within the measurement errors
in both spectra. In contrast, Zeeman splitting has never been reported
for AM CVn systems.

Finally, we beg to disagree with Montgomery et al. (2008) who suggested
that the broad absorption lines seen in a hypothetical carbon analog of
AM CVn could mimic those seen in the photosphere of an isolated white
dwarf. And indeed, it has already been established (see, e.g.,
O'Donoghue \& Kilkenny 1989 or Warner 1995), that the HeI absorption
lines seen in AM CVn systems, although relatively broad by main
sequence standards, have strengths that are more akin to those observed
in log $g$ = 6 hot subdwarf stars than log $g$ = 8 DB atmospheres. This
reflects the fact that they are formed in the inner region of the
accretion disk orbiting the white dwarf primary where the physical
conditions are similar to those encountered in log $g$ = 6
atmospheres. In contrast, the spectral fits carried out by Dufour et
al. (2008b) for Hot DQ stars totally rule out equivalent surface
gravities of log $g$ = 6, as the line strengths observed in these 
stars bear the clear signature of log $g$ = 8 environments, even log $g$
= 9 in the case of SDSS J1426+5752 (notwithstanding the presence of a
magnetic field). In brief, the spectrum of SDSS J1426+5752 cannot be
confused with that of the accretion disk in a hypothetical carbon analog
of AM CVn. 

\section{CONCLUSION}

We have presented an analysis based on follow-up photometric and
spectroscopic observations of the faint but highly interesting star SDSS
J1426+5752. On the photometric front, we carried out a campaign in
integrated light over a baseline spanning some 40 days. We used the
Kuiper/Mont4K combination at the Steward Observatory Mount Bigelow station near
Tucson. Altogether, we acquired some 106.4 h of useful photometry during
the campaign. Our analysis of this data set confirms that the
light curve of SDSS J1426+5752 is dominated by a periodicity at 417.707
s along with its first harmonic (208.853 s) as found originally, but
with less accuracy, by Montgomery et al. (2008) in their discovery
paper. In addition, due to the higher sensitivity achieved in our
campaign, we uncovered the presence of a new oscillation with a period
of 319.720 s, a 4.8 $\sigma$ result using the standard detection
criterion (a 6.0 $\sigma$ detection at the formal level). The
characteristics of these three oscillations are summarized in Table 2.

We investigated the stability of the amplitude and phase of each of the
three periodicities extracted from the light curve of SDSS J1426+5752. Our 
results suggest possible variations over timescales of days, and this is
particularly true for the dominant 417.707 s periodicity. However, this
needs to be confirmed with higher S/N data since observing a $g$ = 19.16
star in integrated light photomery with a small telescope such as the
Kuiper remains challenging. On the basis of our current data, we cannot
be absolutely certain of the reality of the suggested amplitude and
phase modulations in the light curve of SDSS J1426+5752.

We followed up on the suggestion made by Montgomery et al. (2008) that
the luminosity variations in SDSS J1426+5752 may not be caused by
pulsational instabilities, but rather be associted with photometric
activity in a carbon-transferring analog of AM CVn. If true, SDSS
J1426+5752 would represent the prototype of a new class of cataclysmic
variable. Since their argument hinges on the shape of the light curve
folded on the dominant periodicity of 417.707 s, we further exploited
our 106.4 h data set in that direction. Using this and archived light
curves of known pulsating stars, we found, in agreement with Montgomery et
al. (2008), that the folded pulse shape of SDSS J1426+5752 is unusual
compared to those of large amplitude pulsating white dwarfs and hot
subdwarfs. However, in view of the existence of isolated pulsators 
such as the roAp stars HR 3831 and HD 99563 exhibiting pulse shapes with
flatter light maxima than minima (the opposite of what is seen in large
amplitude pulsating white dwarfs), we emphasize that the pulse shape
argument {\sl cannot} be used to discriminate against the pulsations
interpretation. Since SDSS J1426+5752 and these roAp stars and others
share the common property of having a magnetic field sufficiently strong
to affect the pulsations in their atmospheric layers, we suggest instead
that the magnetic field may be responsible for the different pulse shape
as compared to those of (nonmagnetic) pulsating white dwarfs.

On the other hand, arguments against the interacting binary hypothesis
can be put forward. For instance, the light curves we gathered have
shown to be quite regular and stable over at least a six week period, 
a behavior that is not commonly observed in AM CVn systems. Our discovery of a
periodicity (319.720 s) that is not harmonically related to the dominant
oscillation of 417.707 s also goes against the interacting binary
proposal. Likewise, on the spectroscopic front this time, our detailed
radial velocity analysis of the available MMT spectrocopy has revealed
that the velocities are constant to within 7.1 km s$^{-1}$ over a period of
3.37 h. This is to be compared with expected velocity semi-amplitudes in
the range 80$-$100 sin $i$ km s$^{-1}$. Furthermore, the spectrum of
SDSS J1426+5752, unlike those of AM CVn systems, exhibits well defined
symmetric absorption lines and it has proven stable over at least a 20 h
period. It shows sharp Zeeman splitting in the line cores, indicative of
the presence of a large scale magnetic field of 1.2 MG, a property never
observed in a AM CVn system. Finally, the line strengths indicate a
physical environment appropriate for log $g$ = 8 atmospheres, unlike
that found in the inner region of an accretion disk around a white dwarf
and more akin to what is found in a log $g$ = 6 stellar atmosphere.

It should also be pointed out that the interacting binary hypothesis does
not account well at all for the existence of the family of Hot DQ white
dwarfs as a whole since, among other things, not all of them exhibit
luminosity variations (and see Dufour et al. 2009). In contrast, the
single-star evolutionary scenario originally proposed by Dufour et
al. (2007) appears quite viable as demonstrated recently by Althaus et
al. (2009). Also, the nonadiabatic calculations of Fontaine et
al. (2008; also Dufour et al. 2008b) do predict a mixture of pulsators
and nonpulsators in the Hot DQ population. In addition, according to
Montgomery et al. (2008) themselves, citing the work of Benz et
al. (1990), Rasio \& Shapiro (1995), and Piersanti et al. (2003),
carbon-transferring AM CVn-type systems are not expected to exist. 
Putting all these arguments together, we conclude with high confidence
that, of the two possibilities put forward by Montgomery et al. (2008)
to account for the luminosity variations seen in SDSS J1426+5752,
pulsational instabilities is the correct cause. This is also the
preferred solution of Montgomery et al. (2008).

\acknowledgements

This work was supported in part by the NSERC of Canada. P. Dufour is a
CRAQ postdoctoral fellow. G. Fontaine also acknowledges the contribution
of the Canada Research Chair Program.

% References

\end{document}